# Blockchain and its Role in the Internet of Things (IoT)

## Tanweer Alam

Department of Computer Science, Faculty of Computer and Information Systems, Islamic University of Madinah, Saudi Arabia

Email: tanweer03@iu.edu.sa

## ABSTRACT

Blockchain (BC) in the Internet of Things (IoT) is a novel technology that acts with decentralized, distributed, public and real-time ledger to store transactions among IoT nodes. A blockchain is a series of blocks, each block is linked to its previous blocks. Every block has the cryptographic hash code, previous block hash, and its data. The transactions in BC are the basic units that are used to transfer data between IoT nodes. The IoT nodes are different kind of physical but smart devices with embedded sensors, actuators, programs and able to communicate with other IoT nodes. The role of BC in IoT is to provide a procedure to process secured records of data through IoT nodes. BC is a secured technology that can be used publicly and openly. IoT requires this kind of technology to allow secure communication among IoT nodes in heterogeneous environment. The transactions in BC could be traced and explored through anyone who are authenticated to communicate within the IoT. The BC in IoT may help to improve the communication security. In this paper, I explored this approach, its opportunities and challenges.

## I. INTRODUCTION

The IoT is growing exponentially year by year with its aim in 5G technologies, like Smart Homes and Cities, e-Health, distributed intelligence etc. but it has challenges in security and privacy. The IoT devices are connected in a decentralized approach. So, it is very complex to use the standard existing security techniques in the communication among IoT nodes. The BC is a technology the provide the security in transactions among the IoT devices. It provides a decentralize, distribute and publicly available shared ledger to store the data of the blocks that are processed and verified in an IoT network. The data stored in the public ledger is managed automatically by using the Peer-to-peer topology. The BC is a technology where transactions fired in the form of a block in the BC among IoT nodes. The blocks are linked with each other and every device has its previous device address. The blockchain and IoT together work in the framework of IoT and Cloud integration. In the future, the BC would revolutionize the IoT communication [1]. The goals of BC and IoT integration could be summarized as follows.

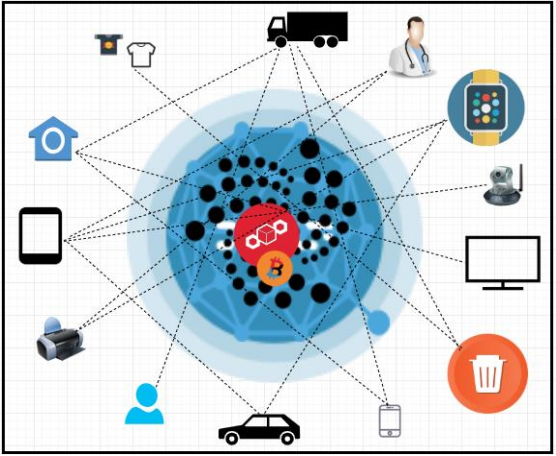

**Figure 1 :** Blockchains and IoT

**i) Decentralized framework:** This approach is similar in IoT and BC. It is removed the centralized system and provide the facility of a decentralized system. It improves the failure probability and performance of the overall system.

**ii) Security:** In the BC, the transactions among nodes are secured. It is a very novel approach for secure communication. BC allows IoT devices to communicate with each other in a secure way.

**iii) Identification:** In IoT, all the connected devices are uniquely identified with a unique ID. Every block in BC is also uniquely identified. So, BC is a trusted technology that provides uniquely identified data stored in public ledger.

**iv) Reliability:** IoT nodes in BC have the capabilities to authenticate the information passed in the network. The data is reliable because it is verified by the miners before entering in BC. Only verified blocks can enter in the BC.

**v) Autonomous:** In BC, all IoT nodes are free to communicate with any node in the network without the centralized system.

**vi) Scalability:** In BC, the IoT devices will communicate in high-available, a distributed

intelligence network that connects with destination device in a real-time and exchange information.

The rest of the paper is summarized as follows: section 1 represents the introduction of the paper, section 2 represents the literature survey, section 3 introduces the role of BC in IoT, section 4 represents the opportunities of the integrated approach, section 5 represents the challenges and section 6 represents the conclusion.

## II. LITERATURE SURVEY

The security and privacy in the communication among IoT devices paid too much attention in the year of 2017 and 2018. Several papers are published during the year 2017 and 2018. In the year of 1990, Stuart Haber and W. Scott Stornetta were written an article [3] on exchanging a document with privacy without storing any information on the time-stamping service. The idea of blockchains comes from [3] but the first blockchains were presented by Satoshi Nakamoto in 2008. He presented a paper where the blocks were added in a chain and form a blockchain [4]. In the article [5], the authors presented the "IoTChain" for authentication of information exchanged between two nodes in an IoT network. They have presented an algorithm to exchange the information in IoT and blockchains (fig 2) [5]. In this paper, authors are focused on the authorization part of the security in the IoTChain framework.

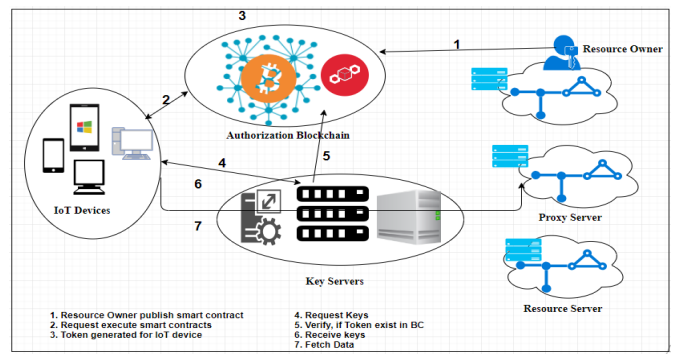

**Figure 2 :** IoT Chain framework

In the article [6], the authors explored the cloud and MANET framework to connect the smart devices in the internet of things and provide communication security. In the article [7], authors represent a very nice framework called internet-cloud framework, it is a good idea to provide secure communication to the IoT devices. In the article [8], the authors provide a middleware framework in the cloud-MANET architecture for accessing data among the IoT devices. Article [9,10] represents the reliability in the communication among IoT nodes. The articles [11,12,13,14,15] are providing the mobility models for communication in 5G networks. In the article [16], the fuzzy logic-based mobility framework is explained for communication security. In the article [17], a nice survey on blockchains and IoT done by the researchers. They present the idea of the security in the BC-IoT to develop the IoT apps with the power of BCs.

### III. THE ROLE OF BC IN IoT

The IoT enables the connected physical things to exchange their information in the heterogeneous network [18]. The IoT could be divided into the following sections.

**1. Physical Things:** The IoT provide the unique id for each connected thing in the network. The physical things are able to exchange data with other IoT nodes.

**2. Gateways:** The gateways are the devices work among physical things and the cloud to ensure that the connection is established and security provided to the network.

**3. Networking:** it is used to control the flow of data and establish the shortest route among the IoT nodes.

**4. Cloud:** It is used to store and compute the data.

The BC is a chain of verified and cryptographic blocks of transactions held by the device connected in a network. The blocks data are stored in the digital ledger that is publicly shared and distributed. The BC provides secure communication in IoT network. The blockchain can be a private, public or consortium with different properties. The following table represents the differentiation among all kind of blockchains.

**Table 1 :** Kinds of Blockchains and their properties

| BC/ Properties | Efficiency | Decentralized | Accord growth | immovableness | Reading | Determining |
|---|---|---|---|---|---|---|
| Private BC | good | No | yes | Can be | Can be publicly | Only one industry |
| Public BC | worse | Yes | no | No | publicly | All miners |
| Consortium BC | good | Some times | yes | Can be | Can be publicly | IoT devices |

The database in blockchains has the properties such as decentralized trust model, high security, highly publicly accessed, privacy is low to high and the transferable identities while in a centralized database, the properties are centralized trust model, low in security, low publicly accessed, privacy is high and non-transferable identities. From the above properties, the blockchain is more advanced than the centralized storage.

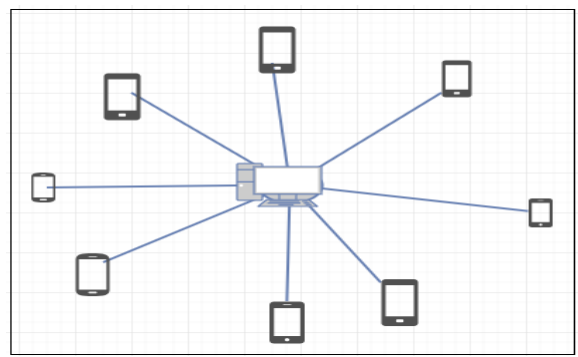

(a)

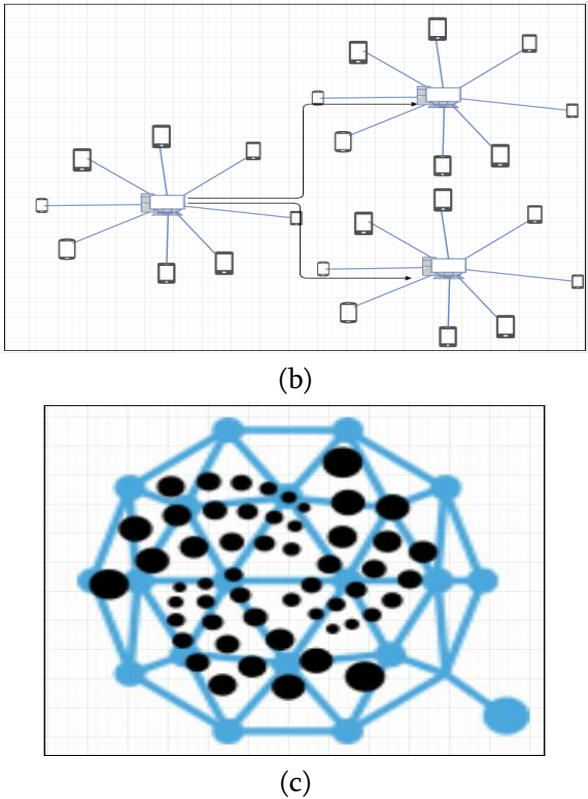

(b)

(c)

**Figure 3 :** (a) Centralized (b) Decentralized (c) Distributed

The following platforms are used to develop IoT applications using blockchain technology.

**a. IOTA:** The IOTA is the new platform for the blockchain and IoT called Next generation blockchains. This platform facilitates the high data integrity, high performance of transactions and high validity of blocks with using fewer resources. It resolves the limitations of blockchains [19].

**b. IOTIFY:** It provides web-based internet of things solution to minimize the limitations of blockchains technology in the form of custom apps [20].

**c. iExec:** It is an open source blockchain based tool. It facilitates your apps the decentralized cloud advantages [21].

**d. Xage:** It is the secure blockchain platform for IoT to increase automation and secure information [22].

**e. SONM:** It is a decentralized blockchain based fog computing platform to provide secure cloud services.

The IoT and blockchains are increasing the business opportunities and opening the new markets where everyone or everything can communicate in a real-time with authenticity, privacy and security in a decentralized approach. The integration of these novel technologies will change the current world where the devices will communicate without the humans in various stages. The objective of the framework is to get the secured data on the right location, on the right format, at real-time. The BC could be used to track billions of IoT connected things, coordinate these things, enabling the processing of the transactions, resolving or eliminating the failures and making the flexible ecosystem for running the physical things on it. Hashing techniques are used in blocks of data by BC to create information privacy for the users.

## IV. OPPORTUNITIES

The BC-IoT integration approach has a lot of remarkable opportunities. It opens the new doors for both together. Some of the opportunities are described as follows.

**1. Building the Trust between parties:** The BC-IoT approach will build trust among the various connected devices because of its security features. Only verified devices can communicate in the network and every block of the transaction will first verify by the miners then they can enter in the BC.

**2. Reduce the Cost:** This approach will reduce the cost because it communicates directly without the third party. It eliminates all the third-party nodes between the sender and the receiver. It provides direct communication.

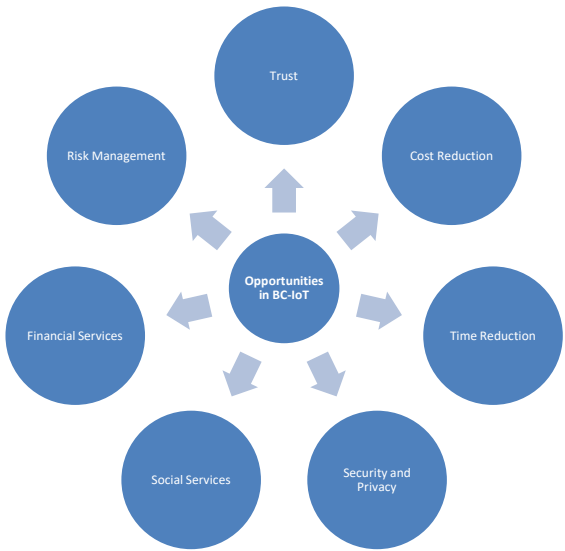

**Figure 4 :** Opportunities in BC-IoT

**3: Reduce Time:** This approach is reduced the time a lot. It reduces the time taken in transactions from days to second.

**4: Security and Privacy:** It provides security and privacy to the devices and information.

**5. Social Services:** This approach provides public and social services to the connected devices. All connected devices can communicate and exchange information between them.

**6. financial Services:** This approach transfer funds in a secure way without the third party. It provides fast, secure and private financial service. It reduced transfer cost and time.

**7. Risk management:** This approach is played the important roles to analyze and reduce the risk of failing the resources and transactions.

## V. CHALLENGES

The IoT and BC could face a lot of challenges such as scale, store, skills, discover etc. The following are the challenges faced by the integration approach.

**1. Scalability:** The BC can become hang because of its heavy load of the transaction. The Bitcoin storage is becoming more than 197 GB storage in 2019 [24].

Imagine if IoT integrates with BC then the load will be heavier than the current situation.

**2. Storage:** The digital ledger will be stored on every IoT node. By the time, it will increase in its storage size that will be a challenging task and become a heavy load on each and every connected device.

**3. Lack of Skills:** The BC is a new technology. It is known by very few people in the world. So, it is also a challenge to train the people about the technology.

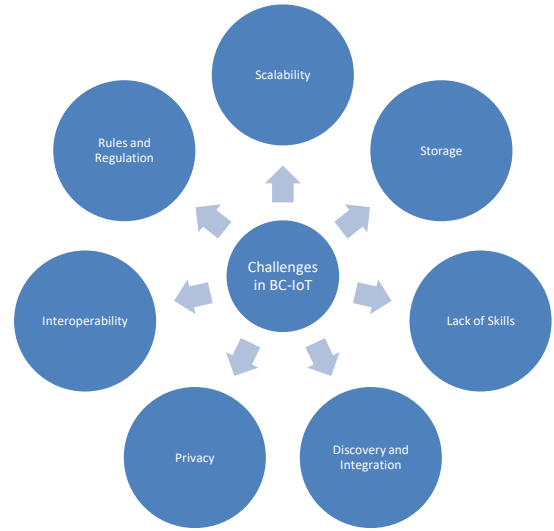

**Figure 5 :** Challenges in BC-IoT

**4. Discovery and Integration:** Actually, BC is not designed for IoT. It is a very challenging task for the connected devices to discover another device in BC and IoT. So, IoT nodes can discover each other but they can be unable to discover and integrate the BC with another device.

**5. Privacy:** The ledger is distributed publicly to every connected node. They can see the ledger transactions. So, privacy is also a challenging task in the integrated approach.

**6. Interoperability:** The BC can be public or private. So, the interoperability between public and private blockchains is also a challenge in the BC-IoT approach.

**7. Rules and Regulation:** The IoT-BC will act globally, so it faces many rules and regulations for implementing this approach globally.

## VI. CONCLUSION

The BC and IoT is a novel approach explored in this article. Many opportunities and challenges are described. Also, available platforms are listed in this article. This approach can be the future of the internet because it can overhaul the current internet system and change it with the new one where every smart device will connect to other devices using the peer-to-peer network in a real-time. It can reduce the current cost and time and provide the right information to the right device in a real-time. So, it can be very useful in the future.

## VII. ACKNOWLEDGMENT

This research is supported by Deanship of Scientific Research, Islamic University of Madinah, Kingdom of Saudi Arabia. The grant number is 10/40.

## Cite this article as :